\documentclass[aps,prl,twocolumn,groupedaddress]{revtex4}
\usepackage{amsmath}
\usepackage{graphicx}
\usepackage{subfigure}
\usepackage{epstopdf}
\usepackage{epsfig}
\usepackage{amssymb}
\usepackage{bm}
\usepackage{bbm}
\newcommand{\beq}{\begin{equation}}
\newcommand{\eeq}{\end{equation}}
\newcommand{\ov}{\overline}

\usepackage{color}

\begin{document}

\title{
 A Resolution of the Flavor Problem of
Two Higgs Doublet Models 
\\ with an Extra $\bm{U(1)_H}$ Symmetry 
for Higgs Flavor
}

\author{P. Ko}
\affiliation{School of Physics, KIAS, Seoul 130-722, Korea}

\author{Yuji Omura}
\affiliation{School of Physics, KIAS, Seoul 130-722, Korea}


\author{Chaehyun Yu}
\affiliation{School of Physics, KIAS, Seoul 130-722, Korea}


\begin{abstract}
\noindent
We propose to implement the Natural Flavor Conservation criterion 
in the two Higgs doublet model (2HDM) to an extra $U(1)_H$ 
gauge symmetry for Higgs flavor, assuming  
two Higgs doublets carry different $U(1)_H$ charges.
Then one can easily avoid the tree-level FCNC from neutral Higgs
mediations using local gauge symmetries, instead of softly broken 
ad hoc $Z_2$ symmetry, and the pseudoscalar boson could be eaten by 
extra $Z_H$ boson. Imposing the anomaly cancellation, we find that 
the $U(1)_H$ in the Type-II and Type-IV 2HDMs become leptophobic 
and leptophilic, respectively. For the Type-I case, $U(1)_H$ depends 
on two parameters,  and some simple cases include   
$U(1)_H = U(1)_{B-L}$, $U(1)_R$, or $U(1)_Y$.
We sketch qualitative phenomenology of these models.
\end{abstract}


\maketitle

\section*{Introduction\label{sec:intro}}

The Higgs sector is the least understood part of the standard model (SM), 
both theoretically and experimentally. It is anticipated that the LHC will
probe the SM Higgs sector fully within coming years, and provide us with 
invaluable informations about the origin of electroweak (EW) symmetry 
breaking.  Being least understood, one could consider 
various extensions of the SM Higgs sector.
Adding one more Higgs doublet to the SM is one of the simplest 
extensions of the SM, the two Higgs doublet model (2HDM). 
The 2HDMs have been studied in various contexts (see, for example,
Ref.~\cite{Branco:2011iw} for a recent review).

Generic 2HDMs suffer from excessive flavor changing neutral current 
(FCNC) mediated by neutral Higgs boson exchanges. 
One way to avoid this problem is to impose an ad hoc $Z_2$ discrete 
symmetry as suggested by Glashow and Weinberg long time ago 
\cite{Glashow:1976nt}, 
which is often called Natural Flavor Conservation (NFC): 
\[
Z_2 : ( H_1 , H_2 ) \rightarrow (+H_1 , - H_2 ) .
\]
The Yukawa sectors can be controlled by assigning suitable $Z_2$
parities to the SM fermions, and the models are often categorized 
into four types as shown in Table~\ref{table1}~\cite{Barger:1989fj,Aoki:2009ha}.
However it is well known that discrete symmetry could generate a 
domain wall problem when it is spontaneously broken, which is indeed
the case in the 2HDM. Therefore the $Z_2$ symmetry is assumed to be 
broken softly by a dim-2 operator, $H_1^\dagger H_2$ term. 
Also the origin of such a discrete symmetry is not clear at all. 
\begin{table}[th]
\begin{center}
\caption{
\label{table1}%
{
Assignment of $Z_2$ parities to the SM fermions and Higgs doublets.
}
}
\begin{tabular}{|c||cc|cc|cc|c|}
\hline 
Type & $H_1$ & $H_2$ & $U_R$ & $D_R$ & $E_R$ & $N_R$ & $Q_L, L$
 \\
 \hline
  I  & $+$   & $-$   & $+$   & $+$   & $+$   & $+$   & $+$ 
 \\
 II  & $+$   & $-$   & $+$   & $-$   & $-$   & $+$   & $+$
 \\
 III & $+$   & $-$   & $+$   & $+$   & $-$   & $-$   & $+$
 \\
 IV  & $+$   & $-$   & $+$   & $-$   & $+$   & $-$   & $+$    
 \\ \hline
\end{tabular}
\end{center}
\end{table}

In this Letter, we propose to replace the $Z_2$ symmetry in 2HDM 
by a new $U(1)_H$ symmetry associated with Higgs flavors, 
where $H_1$ and $H_2$ have different $U(1)_H$  charges
and thus are distinguished by new $U(1)_H$ gauge boson $Z_H$. 
The SM fermions can be either charged or neutral under $U(1)_H$.
Then the NFC criterion suggested by Glashow and Weinberg has
its origin in the framework of local gauge symmetry $U(1)_H$ 
for Higgs flavor.
The generic feature of our models would be that there should be 
an extra spin-1 gauge boson $Z_H$.  Its couplings to the SM 
fermions are completely controlled by $U(1)_H$ charges of the SM 
fermions and two Higgs fields, with phenomenologically acceptable
Yukawa interactions and anomaly cancellation. 
Our models have qualitatively different aspects from the usual 2HDM, 
and thus deserve more close study both at colliders and 
at low energy flavor physics as well as in the context of 
electroweak phase transition and baryogenesis. 
In this Letter, we will present the definitions of the models and describe 
the basic features (mostly at tree level), relegating more detailed 
comprehensive analysis for future study. 

\section*{Higgs sector}

Let us assume that $H_1$ and $H_2$ carry different $U(1)_H$ charges, 
$h_1$ and $h_2$ (with $h_1 \neq h_2$ in order to distinguish two of them), 
with $g_H$ being the $U(1)_H$ coupling.  The kinetic terms for the $H_1$ and 
$H_2$ will involve the $U(1)_H$ couplings:
\begin{equation}
D_\mu H_i = D_\mu^{\rm SM} H_i - i g_H h_i Z_{H \mu} H_i
\end{equation}
with $i=1,2$. 
Then the mass matrix for $Z$ and $Z_H$ from the kinetic terms of
$H_1$ and $H_2$ is given by 
\begin{equation}
M^2 = \left( \begin{array}{cc}
                    g_Z^2 v^2   &   -g_Z g_H ( h_1 v_1^2 + h_2 v_2^2 )  \\
                    -g_Z g_H ( h_1 v_1^2 + h_2 v_2^2 )  & 
                    g_H^2 ( h_1^2 v_1^2 + h_2^2 v_2^2 ) 
                    \end{array}   \right) \ ,
\end{equation}
where $v^2 = v_1^2 + v_2^2$. 
Note that the determinant of $M^2$ is not zero, as long as 
$h_1 \neq h_2$.
If we add an additional $U(1)_H$ charged singlet scalar $\Phi$ 
(its $U(1)_H$ charge is defined as $h_\phi$) with nonzero 
vacuum expectation value (VEV) 
$v_\phi$,  the $(22)$ component of the (mass)$^2$ matrix 
would have an additional piece  $g_H^2 h^2_{\Phi} v_\Phi^2$ from the 
kinetic term of $\Phi$.  The mass mixing must be small to avoid too large 
deviation of $\rho$ parameter  from the SM prediction. 
The tree-level deviation within $1\sigma$ 
restricts the mass and coupling of $Z_H$:
\begin{equation}
\label{eq:Constraint-Rho}
 \{ h_1 (\cos \beta )^2+ h_2 (\sin \beta )^2 \}^2
 \frac{ g^2_H}{ g^2_Z}\frac{  m_{\Hat{Z}}^2 }{ m_{\hat{Z}_H}^2 - m_{\Hat{Z}}^2}
 \lesssim O(10^{-3}),
\end{equation}
where $ m_{\Hat{Z}}^2 = g_Z^2 v^2$ and 
$ m_{\hat{Z}_H}^2= g_H^2  v^2 \{ h^2_1 (\cos \beta )^2+h^2_2 (\sin \beta )^2 \}
+g_H^2  h^2_{\phi} v_\phi^2  $.

The potential of our 2HDM is given by 
\begin{eqnarray}
V (H_1,H_2) & = & m_1^2 H_1^\dagger H_1 + m_2^2 H_2^\dagger H_2 
+ \frac{\lambda_1}{2} ( H_1^\dagger H_1 )^2 
\nonumber \\
&& \hspace{-13ex}
+ \frac{\lambda_2}{2} ( H_2^\dagger H_2 )^2 
+ \lambda_3 H_1^\dagger H_1 H_2^\dagger H_2 
+ \lambda_4 H_1^\dagger H_2 H_2^\dagger H_1 .
\end{eqnarray}
In terms of the standard notation for the 2HDM potential, our 
model corresponds to a special case $m_3^2 = \lambda_5 = 0$. 
Note that $H_1^\dagger H_2$ or its square are forbidden by $U(1)_H$ 
symmetry, since we have imposed $h_1 \neq h_2$.
If the model were not gauged with the extra $U(1)_H$, one would
encounter the usual problem of a massless pseudoscalar $A$.
In our case, this massless mode is eaten by the $U(1)_H$ gauge boson,
and there is no usual problem with a massless Goldstone boson.
Instead the scalar boson spectrum is different from the usual 
2HDM, since there would no pseudoscalar $A$ in our models.

In case we include a singlet scalar $\Phi$, let us define $\phi = h_1 - h_2$, 
so that $H_1^\dagger H_2 \Phi$ is gauge invariant.  
Then there would be additional terms in the scalar potential:
\begin{eqnarray}
\Delta V &=& m_\Phi^2 \Phi^\dagger \Phi
+ \frac{\lambda_\Phi}{2} (\Phi^\dagger \Phi )^2
+ ( \mu H_1^\dagger H_2 \Phi + \textrm{h.c.} ) 
\nonumber \\
&+& \mu_1 H_1^\dagger H_1 \Phi^\dagger \Phi
+ \mu_2 H_2^\dagger H_2 \Phi^\dagger \Phi , 
\end{eqnarray}
depending on $h_1$, $h_2$ and $h_\phi$. 
After $\Phi$ develops a VEV, $\mu$ terms look like the 
$m_3^2$ term in the conventional notation. And the effective 
$\lambda_5$ term is generated by the $\Phi$ mediation: 
$\lambda_5 \sim \mu^2 / m_\Phi^2$ well below $m_\Phi$ scale.
In any case there is no dangerous Peccei-Quinn symmetry leading 
to a massless $Z^0$ unlike the usual 2HDM, and no need for
soft breaking of $Z_2$ symmetry, because of extra $U(1)_H$ 
gauge symmetry.


Production and decay modes of the new $Z_H$ gauge boson will depend on 
the $U(1)_H$ charges of the SM fermions, which will differ case by case.
In the following, we implement each 2HDM  
with NFC (Type-I, II, III, IV) to local 
$U(1)_H$ gauge theories by assigning suitable $U(1)_H$  charges to two Higgs
doublets $H_1$ and $H_2$ and the SM fermions, and by adding new chiral 
fermions for anomaly cancellation.

\section*{Type-I 2HDM\label{sec:type1}}

Let us first start with the simplest case, the Type-I 2HDM, where 
the SM fermions can get masses only from $H_1$ VEV. 
This is possible, if (with $h_1 \neq h_2$)
\begin{equation}
u - q  - h_1 = d - q + h_1 = e - l + h_1 = n - l - h_1 = 0 .
\end{equation}
There are many ways to assign $U(1)_H$ charges to the SM fermions
to achieve this scenario.  The phenomenology will depend crucially 
on the $U(1)_H$ charge  assignments of the SM fermions. 
In general, the models will be anomalous, even if $U(1)_H$ charge 
assignments are nonchiral, so that one has to achieve anomaly 
cancellation by adding new chiral fermions to the particle spectrum. 

\begin{table}[th]
\begin{center}
\caption{
\label{table2}%
{
Charge assignments of an anomaly-free $U(1)_H$ in the Type-I 2HDM.
}
}
\begin{tabular}{|c|cc|ccccc|}
\hline 
Type & $U_R$ & $D_R$ & $Q_L$ & $L$ & $E_R$ & $N_R$ & $H_1$ 
\\
\hline
$U(1)_H$ charge & $u$   & $d$   & $\frac{(u+d)}{2}$ & $\frac{-3 (u+d)}{2}$ & 
$-(2 u + d)$ & $-(u+2d)$ & $ \frac{(u-d)}{2}$
\\
\hline
$h_2 \neq 0$ & $0$   & $0$    & $0$   & $0$    & $0$  & $0$  & $0$  
\\
$U(1)_{B-L}$ & $1/3$ & $1/3$  & $1/3$ & $-1$   & $-1$ & $-1$ & $0$ 
\\
$U(1)_R$  & $1$   & $-1$   & $0$   & $0$    & $-1$ & $1$  & $1$  
\\
$U(1)_Y$ & $2/3$ & $-1/3$ & $1/6$ & $-1/2$ & $-1$ & $0$  & $ 1/2$
 \\ \hline
\end{tabular}
\end{center}
\end{table}
For the Type-I case, one can achieve an anomaly-free $U(1)_H$ 
assignment even without additional chiral fermions as in Table~\ref{table2}. 
There is one free parameter by which the charge assignments determines
the theory, modulo the overall coupling constant $g_H$. It is amusing 
to observe that there appear an infinite number of new models which is 
a generalization of the Type-I model into Higgs flavor $U(1)_H$ models 
without extending the fermion contents at all. 

There are four simple and interesting anomaly-free charge assignments 
without new chiral fermions, however: 
\begin{itemize}
\item $(u,d) = (0,0)$: 
In this case, all the SM fermions are $U(1)_H$ singlets.
Then $Z_H$ is fermiophobic and Higgsphilic. It would not be easy
to find it at colliders because of this nature of $Z_H$, and $h_2 \neq 0$. 
In this case, $H^\pm W^\mp Z_H$ couplings from the Higgs kinetic 
terms would be the main source of production and discovery for $Z_H$. 
The phenomenology of $Z_H$ will be similar to the leptophobic 
$Z'$ studied in Ref.~\cite{Georgi:1996ei}.
\item $(u,d) = (\frac{1}{3}, \frac{1}{3})$:
In this case, we have $U(1)_H = U(1)_{B-L}$, and 
$Z_H$ is the $(B-L)$ gauge boson, which gets 
mass from the doublet $H_2$ (and also by a singlet $\Phi$, if 
we include it).
Our case is very different from the usual $(B-L)$ model where 
$U(1)_{B-L}$ is broken only by the SM singlet scalar $\Phi$.
Therefore the phenomenology would be very different. 
However the Yukawa sector is controlled by $U(1)_H$ and a new 
Higgs doublet $H_2$ with nonzero $U(1)_H$ charge $h_2$.  
\item $(u,d) = (1, -1)$:
In this case, we have $U(1)_H = U(1)_R$. 
The $Z_H$ couples only to the right-handed (RH) fermions, 
not to the left-handed (LH) fermions.
In this case, the would-be SM Higgs doublet $H_1$ also carries 
nonzero $U(1)_H$ charge, and Higgs phenomenology of this
type of models will be very different from the SM Higgs boson.
\item $(u,d) =  (\frac{2}{3}, -\frac{1}{3})$: 
This case corresponds to $U(1)_H = U(1)_Y$. 
If $h_2 \neq h_1$ is satisfied, $H_2$ does not couple 
with the SM fermion at tree level,
and this model becomes different from the SM.
In this case, $H_1$ can couple to the $Z_H$ boson and $H_2$ directly
while the SM fermions couple to the $Z_H$ boson directly
and eventually to $H_2$ through a $Z_H$ loop. These interactions will change
phenomenology of this type of models from the SM Higgs boson. 
\end{itemize}
Other interesting possibilities with vector-like $U(1)_H$ are
to identify $U(1)_H = U(1)_B$ or $U(1)_L$. In these cases, 
however, the model becomes anomalous, and we have add 
additional chiral fermions for anomaly cancellation. 
Again, it is interesting to break $U(1)_B$ or $U(1)_L$ by
an $SU(2)_L$ doublet $H_2$ (and possibly by $\Phi$ too).


\section*{Type-II and Type-IV 2HDMs \label{sec:type2}}

In this section, we will implement the Type-II model to a 
$U(1)_H$ gauge theory.
In the Type-II 2HDM, $H_1$ couples to the up-type fermions, 
while $H_2$ couples to the down-type fermions: 
\begin{eqnarray}\label{eq;type2}
V_y &=& y_{ij}^U \overline{Q_{Li}} \widetilde{H_1} U_{Rj}
      + y_{ij}^D \overline{Q_{Li}} H_2 D_{Rj}
\nonumber \\ 
    &+& y_{ij}^E \overline{L_i} H_2 E_{Rj}
      + y_{ij}^N \overline{L_i} \widetilde{H_1} N_{Rj}+\textrm{h.c.}
\end{eqnarray}
For example, 
we can consider the $U(1)_H$ charge assignment, that $u=n=h_1$ are 
only nonzero, in order to get the Type-II Yukawa couplings.

\begin{table}[ht]
\begin{center}
\caption{
\label{table3}%
{
Charge assignments to the extra chiral fields in the Type-II case.
}
}
\begin{tabular}{|c|c|c|c|c|c|c|}\hline
         & $SU(3)$ & $SU(2)$ & $U(1)_Y$ & $U(1)_H$ \\ \hline 
$q_{Li}$ & $3$     & $1$     & $2/3$    & $\Hat{Q}_L=u+\Hat{Q}_R$ \\ \hline 
$q_{Ri}$ & $3$     & $1$     & $2/3$    & $\Hat{Q}_R$      \\ \hline 
$n_{Li}$ & $1$     &$1$      & $0$      & $\Hat{n}_L=u+\Hat{n}_R$ \\ \hline 
$n_{Ri}$ & $1$     &$1$      & $0$      & $\Hat{n}_R$ \\ \hline      
\end{tabular}
\end{center}
\end{table}
There could be a number of ways to achieve anomaly cancellation, and
one simple way is to add SM vector-like pairs. 
The $U(1)_H$ charge assignment to the extra chiral fields can be given
like in Table~\ref{table3}. Here, we require 
$3  \Hat{Q}_R \Hat{Q}_L +  \Hat{n}_R \Hat{n}_L =0$ for $U(1)_H^3$ 
anomaly cancellation.  
The $U(1)_YU(1)_H^2$ anomaly is nonzero with this matter content, 
so that we need to add more SM vector-like pairs,
which are introduced in Refs.~\cite{Ko:2011vd,Ko:2011di}. 
For example, two pairs of $SU(2)_L$ doublets 
with $-1/2$ $U(1)_Y$, $(l'_{L1,2},l'_{R1,2})$, 
or $SU(3)$ triplets with $-1/3$ $U(1)_Y$, $(q'_{L1,2},q'_{R1,2})$, 
can cancel the $U(1)_YU(1)_H^2$ anomaly
without disturbing other conditions. Let us define the charges 
of $(l'_{L1},l'_{R1})$ (or $(q'_{L1},q'_{R1})$) as  $(Q_L,Q_R)$, 
and the charges of $(l'_{L2},l'_{R2})$ (or $(q'_{L2},q'_{R2})$) 
as $(-Q_L,-Q_R)$. 
Then, the above charge assignment requires 
$(Q_L,Q_R)=(u/2 +6 \Hat{Q}_R,-u/2 +6 \Hat{Q}_R)$.
The mass terms are given by the $\Phi$ with $U(1)_H$ charge, $u$,
\begin{eqnarray} \label{eq;type2-extra}
V_m &=&  y_{ij}^q \overline{q_{Li}} q_{Rj} \Phi 
      +  y_{ij}^n \overline{n_{Li}} n_{Rj} \Phi  
\nonumber \\
    &+&  y_{1}^l \overline{l'_{L1}} l'_{R1} \Phi 
      +  y_{2}^l \overline{l'_{L2}} l'_{R2} \Phi^{\dagger}+\textrm{h.c.} 
\end{eqnarray}
$\Phi$ should be added for the nonzero masses of the extra fields and
in order to forbid the mixing terms which cause FCNC, such as $m_{ij}\overline{q_{Li}} {U_{Rj}}$ and $\lambda_{ij} \Phi \overline{q_{Li}} {U_{Rj}}$,
we have to require the condition for the charges, $\Hat{Q}_R, Q_R, \Hat{n}_R \neq 0, \pm u$.
On the other hand, without the mixing terms between the SM fermions and $ q_{L,Ri}$,
the extra quarks would be stable without further extensions. 
For example, let us add a SM-singlet scalar, $X$, with $U(1)_H$ charge, $\Hat{Q}_R$. 
$X$ can have the Yukawa couplings, $\lambda_{ij} X \overline{q_{Li}} U_{Rj}$, which allow the extra quark 
to decay.
We can avoid tree-level FCNC contributions if $X$ does not get VEV, and 
$X$ becomes a cold dark matter (CDM) candidate \cite{Ko:2011vd, Ko:2011di}. 
$l'_{L,R1}$ and $l'_{L,R2}$ is also stable before EW symmetry breaking,
but the masses of charged and neural are split according 
to the radiative correction, and then the neutral components
also become good CDM candidates. If $q'_{L,R}$ are adopted instead 
of $l'_{L,R}$, another scalar may be required to destabilize $q'_{L,R}$. 

\begin{table}[ht]
\begin{center}
\caption{
\label{table4}%
{
Charge assignments to the extra chiral fields in the leptophobic $Z^\prime$\
model
in the context of $E_6$~\cite{Rosner:1996eb}.
}
}
\begin{tabular}{|c|c|c|c|c|c|c|}\hline
         & $SU(3)$ & $SU(2)$ & $U(1)_Y$ & $U(1)_H$ \\ \hline 
$q_{Li}$ & $3$     & $1$     & $-1/3$   & $2/3$    \\ \hline 
$q_{Ri}$ & $3$     & $1$     & $-1/3$   & $-1/3$   \\ \hline 
$l_{Li}$ & $1$     & $2$     & $-1/2$   & $0$      \\ \hline 
$l_{Ri}$ & $1$     & $2$     & $-1/2$   & $-1$     \\ \hline   
$n_{Li}$ & $1$     & $1$     & $0$      & $-1$     \\ \hline      
\end{tabular}
\end{center}
\end{table}
$(q,u,d)=(-1/3,2/3,-1/3)$ and $(l,e,n)=(0,0,1)$ case corresponds to 
the leptophobic $Z^{'}$ model in the context of $E_6$ \cite{Rosner:1996eb},
with the following identification of $U(1)_H$ charge in terms of $U(1)$ 
generators of $E_6$ model. 
\[
Q_H = I_{3R} - Y_L + \frac{1}{2} Y_R .
\] 
The extra chiral fields introduced in Ref.~\cite{Rosner:1996eb} 
cancel the anomaly.
The charge assignment to the extra chiral fields is shown 
in Table~\ref{table4}.
The qualitative predictions made in Ref.~\cite{Rosner:1996eb} will 
apply in our case too without any modification.
The Yukawa couplings for SM fermions are given by the Eq.~(\ref{eq;type2}),
and the mass and mixing terms of the extra fermions will be generalized to 
\begin{eqnarray}
  V_m &=& y_{ij}^n \overline{n_{Li}} H_2 l_{Rj}
        + y_{ij}^q \overline{q_{Li}} q_{Rj} \Phi
        + y_{ij}^l \overline{l_{Li}} l_{Rj} \Phi
\nonumber \\
      &+& Y_{ij}^q \overline{Q_{Li}} H_2 q_{Rj} 
        + Y_{ij}^E \overline{l_{Li}} H_2 E_{Rj}
        + Y_{ij}^N \overline{l_{Li}} \widetilde{H_1} N_{Rj} 
\nonumber \\
      &+& Y_{ij}^D \overline{q_{Li}} D_{Rj} \Phi
        + Y_{ij}^l \overline{L_i} l_{Rj} \Phi + \textrm{h.c.} 
\end{eqnarray}
Under this charge assignment, corresponding to $E_6$, the mixing terms 
between the SM fermions and the extra fermions are allowed at tree level, 
so that their Yukawa coupling must be tuned to avoid
the strong constraints from FCNC processes.  
 
The Type-IV model,
where $H_1$ couples to the SM up-type quarks and charged leptons
and $H_2$ couples to the SM down-type quarks and neutral leptons,
could be realized by nonzero $u =e \neq 0$. 
This model is corresponding to the flipped  Type-II,
and the matter set, which was introduced here,
could be applied to this model, by defining $(n_{Li},n_{Rj})$ 
as the fields with $-1$ $U(1)_Y$ charge corresponding to $E_R$. 
$(Q_L,Q_R)$ should be $(u/2+6\Hat{Q}_R-3\Hat{n}_R,-u/2+6\Hat{Q}_R-3\Hat{n}_R)$.

\section*{Type-III 2HDM}
Type-III 2HDM may be called leptophobic or leptophilic models, because 
the $U(1)_H$ charge assignment where only quarks or leptons are charged can
realize the Yukawa couplings of Type-III 2HDM,
\begin{eqnarray}\label{eq;type3}
V_y&=& y^u_{ij} \ov{Q_{Li}  } \widetilde{H_1} U_{Rj} 
      +y^d_{ij} \ov{Q_{Li}  } H_1 D_{Rj}              \nonumber
 \\ 
   &+&y^L_{ij} \ov{L_i}H_2 E_{Rj}+ y^N_{ij} \ov{L_i} \widetilde{H_2} N_{Rj} 
+\textrm{h.c.}
\end{eqnarray}    
The Yukawa couplings require $u - q  - h_1 = d - q + h_1$ and $e - l + h_2 = n - l - h_2.$
For example, $q=\frac{u+d}{2}$, $h_1=\frac{u-d}{2}$ and $e=l=n=h_2=0$ 
leads leptophobic $U(1)_H$ symmetry. 
$u+d=0$ may also be required for $U(1)_H^3$.
The set similar to the Type-II model could cancel the anomaly, 
changing the charge assignment of the extra fields properly,
and good cold dark matter candidates could be realized, 
as mentioned in Refs.~\cite{Ko:2011vd,Ko:2011di}.

Similarly, leptophilic $U(1)_H$ could be also discussed 
according to the assignment, $q=u=d=h_1=0$ and $e=-n$.  
Furthermore, we may consider $e=0$ and $n \neq 0$ cases.
In this case, RH neutrino is only charged and 
we could classify it as Type-III$_n$, where $H_2$ couples only RH neutrino.  
In this case, only $n_{L,Ri}$, which introduced in the Type-II, 
are required for anomaly conditions, 
and the $U(1)_H$ charges, $(\Hat{n}_L, \Hat{n}_R)$, 
should be $(0,n)$ or $(n,0)$ because of $U(1)_H^3$ condition.
In principle, we can also discuss the case that only the RH lepton 
is charged ($e \neq 0$ and $n=0$).

Furthermore, we can also consider the model where only RH up-type 
(down-type) quarks couple with $H_2$, and the RH down-type (up-type) quarks 
and both leptons couple with $H_1$. In that case, only RH up-type (down-type) 
quarks and $H_2$ are charged under $U(1)_H$.
The anomaly condition could be discussed in the same ways of Type-II.

\section*{Phenomenology}

The 2HDMs we newly propose in this work are an extension of 
the usual 2HDMs where the usual discrete $Z_2$ symmetry is implemented
to local gauge symmetry $U(1)_H$.  Thus, phenomenology of our models
is similar to that of the usual 2HDMs, except that there appear new particles  
associated with $U(1)_H$ gauge symmetry and its breaking 
(a new gauge boson $Z_H$ and extra neutral scalar boson from $\Phi$)  
and extra chiral fermions (depending on the types of 2HDMs, 
$U(1)_H$ charge assignment, and how to achieve gauge-anomaly cancellation).
Some of these extra new particles might have influences on the production 
rates and decay branching ratios of  Higgs bosons. 
In certain models, we find that some particles are stable and 
can play a role of nonbaryonic cold dark matter of the universe~
\cite{Ko:2011vd,Ko:2011di}. 

Recently a SM Higgs-like scalar boson with mass $\sim 125$ GeV 
has been observed by CMS and ATLAS Collaborations~\cite{higgs}. 
If this new particle is identified as a Higgs boson in the future, 
it could be regarded as the lightest Higgs boson $h$ in our models.
In general, the extra fermions would reduce the branching ratio 
of the Higgs boson decay into two photons for a $125$ GeV Higgs boson, 
if there is no mixing between extra fermions. On the other hand,
the extra fermions may enhance the Higgs production rate in $gg$ fusion.
However, if there is a large mixing between extra fermions, the contribution 
to $h\to \gamma \gamma$ and $gg\to h$ could be opposite
to the case of no mixing~\cite{Carena:2012xa}.  

The mixing between the SM singlet $\Phi$ and the neutral CP-even 
Higgs bosons would also change the production rates and the branching 
ratios of neutral Higgs bosons. 
The neutral Higgs bosons may also decay into a pair of CDM,  
if the CDM is lighter than the half of the Higgs boson masses.
The current Higgs search at the LHC and Tevatron  gives strong
constraints on the heavier Higgs boson $H$, which are controlled by
the heavier Higgs boson mass $m_H$ and its branching ratios to the 
non-SM particles. These issues are more model-dependent, and 
more careful study on the Higgs boson phenomenology is required 
for each type of 2HDMs proposed in this Letter.   
This is left for a future work~\cite{progress}.

Search for charged Higgs bosons and CP-odd Higgs boson would 
depend on the detail of models too.
In each type of 2HDMs, phenomenology of those Higgs bosons
has been widely discussed in Refs.~\cite{Aoki:2009ha,Yagyu:2012qp}.
The charged Higgs boson mass is stringently constrained by $B$ decays.
The most stringent bounds on the Type-II and Type-IV 2HDMs 
come from the radiative process $b\to s \gamma$~\cite{Borzumati:1998tg}, 
which is about 300 GeV~\cite{Misiak:2006zs}.
We note that the charged Higgs bosons in the Type-I and Type-III
2HDMs are loosely constrained by the $b\to s \gamma$ process. 
The decay $B\to \tau\nu$ can also provide useful constraints 
on the charged Higgs boson in the Type-II~\cite{Hou:1992sy}. 
However, the present constraints are weaker  than that from the 
$b\to s \gamma$ process except for the large $\tan \beta$ region
~\cite{Aoki:2009ha}.
Another interesting processes which may constrain the charged Higgs 
boson are $B\to D^{(\ast)} \tau \nu$ decays. 
Combination of their branching ratios deviates 
from the SM prediction by $3.4 \sigma$~\cite{Lees:2012xj}.
An interesting point is that it is more difficult to accommodate 
the branching fractions with 2HDM of any type with NFC. 
If the  $B\to D^{(\ast)} \tau \nu$ branching ratios settle down 
at the present values, all the 2HDMs with NFC would be ruled out and
one may have to rely on nonminimal flavor violation in 2HDMs
~\cite{Crivellin:2012ye}.

Direct collider bounds for the charged Higgs boson $H^\pm$ come 
from the exotic top quark decay in the light mass region. 
For heavier $H^\pm$ mass, collider bounds come from the direct 
production in $g g/q\bar{q}\to t\bar{b}H^-$, $g b\to t H^-$,
$g g \to H^+ H^-$, {\rm etc}.
The CP-odd neutral Higgs boson $A$ may be searched 
for through the direct productions in 
$g g \to A$, $q\bar{q} \to q^\prime\bar{q^\prime}A$, 
$q\bar{q}\to t\bar{t} A$, {\rm etc}. There is no significant bound
on $H^\pm$ and $A$ from direct search in the high mass region.

Our models also predict a lot of new collider signatures, 
involving $Z_H$ and the extra matters that were introduced 
for anomaly cancellation.  
Models with extra fermions become model-dependent, 
and we do not discuss this sector in detail here.  
Once $Z_H$ couples with leptons, $g_H$ and $m_{Z_H}$ will be 
strongly constrained by the LEP and Drell-Yang processes: 
$m_{Z_H}/g_H \gtrsim 6$ TeV is required in 
$U(1)_{B-L}$ model~\cite{Carena:2004xs}.

If $U(1)_H$ charges are assigned to only quarks and Higgs fields 
and the mixing between $Z_H$ and the SM gauge bosons is small,
the bound would be relaxed drastically and the $Z_H$ interaction could be 
compatible with the $W$ and $Z$ boson interactions. 
This case corresponds to Type-III 2HDM, and $U(1)_B$ model in Type-I,
when we add  extra matters for anomaly cancellation. 
In fact, the $U(1)_H \equiv U(1)_B$ model has been discussed 
in the small mass range of $Z_H$ in Ref.~\cite{wise}.
Around $m_{Z_H}=160$ GeV, the dijet search at Tevatron does not provide 
useful constraints because of the large background. 
Therefore $g_H$ could reach around $0.2$ which corresponds to 
the upper bound from the UA2 experiment~\cite{ua2}.  
If the $Z_H$ mass is around $O(10)$ GeV, the hadronic decays of $Z$ and 
$\Upsilon$ constrain $g_H$ and $m_{Z_H}$~\cite{Carone:1994aa}. 
The mass mixing between $Z$ and $Z_H$ would break the leptophobic behavior
of $Z_H$.  If one wants to keep the leptophobic nature of $Z_H$ for 
some phenomenological motivations,  one has to tune $\tan \beta$ 
in order to satisfy Eq.~(\ref{eq:Constraint-Rho}), 
which is the condition for the small mixing case.

Among the extra matters, there are good CDM candidates which interact 
with the SM quarks through $Z_H$ exchange. Such CDMs which are charged 
under $U(1)_H$ are considered in Ref.~\cite{Ko:2011ns} and the mono-jet 
and mono-photon signal, 
$pp \rightarrow j (\gamma)+Z_H \rightarrow j (\gamma)+{\rm CDMs}$, 
could be one of the best channels to test our models at the LHC~\cite{monojet}. 
  
In the so-called fermiophobic case, which belongs to the Type-I with the SM 
fermions being $U(1)_H$ neutral, the new gauge boson $Z_H$ interacts 
with only Higgs fields in the limit that the mass mixing and kinetic mixing 
with $Z$ are zero.  In this case, $Z_H$ could be light and there is no extra 
fermions or dark matter candidates. 
If $Z_H$ is lighter than the lightest Higgs boson $h$, the lightest Higgs boson 
can decay through $h \to Z_H^{(\ast)} Z_H^{(\ast)} \ , \ Z_H Z^*$, 
followed by $Z_H$ decays to the SM particles though the small mixing with $Z$ boson.  
Or gauge bosons can also decay to $Z_H$ and the off-shell lightest Higgs boson. 
The searches for the exotic decays of the Higgs bosons and $Z$ boson 
are also efficient to find the extra gauge boson $Z_H$.

\section*{Conclusion}

Let us summarize our results. 
In this Letter, we proposed a new resolution of the Higgs-mediated FCNC
problems in 2HDMs using a new $U(1)_H$ gauge symmetry. If two Higgs 
doublets carry different charges under this new $U(1)_H$ symmetry 
and the SM fermions charges are assigned properly, one can easily 
realize the ``Natural Flavor Conservation" suggested by Glashow and 
Weinberg \cite{Glashow:1976nt}.   There are infinitely many ways to 
assign $U(1)_H$ charges compatible with NFC, unlike the common 
practice based on discrete $Z_2$ symmetries.  Our proposal for Type-II 
2HDM has vastly different consequences from the MSSM 2HDM. In the
MSSM, the supersymmetric parts of the Higgs potential is Type-II, 
but eventually all the fermions couple to both Higgs fields
when the loop corrections involving trilinear couplings are included.  
Then Higgs-mediated flavor violation can be enhanced by significant amount, 
especially for the large $\tan\beta$ region in the MSSM case. 
On the other hand, our models for Type-II are based on $U(1)_H$ gauge 
symmetry which is spontaneously broken.  The Higgs mediated FCNC may not be
enhanced much even if we include the loop effects, unlike the MSSM.
We believe our proposal newly opens a wide window for the 2HDMs.  
The Higgs-mediated FCNC problem is no longer a serious problem.  
Instead of pseudoscalar $A$, we have a new massive gauge boson $Z_H$. 
Even for the Type-I case without extra chiral fermions, the phenomenology 
can be very rich  since $Z_H$ can couple to $(B-L)$ of the SM fermions or 
purely fermiophobic, or only to the RH fermions, to name only a few possibilities.  

Phenomenology of our models depend on the details of our models: 
type, $U(1)_H$ charge assignments to the SM fermions and two Higgs doublets
$H_1$ and $H_2$, and extra chiral fields, etc.
In general, our models have stringent constraints from search for
$Z^\prime$ boson,  neutral and charged Higgs bosons and extra fermions. 
Strong indirect bounds on charged Higgs boson and $\tan\beta$ come
from  radiative and (semi)leptonic $B$ decays. 
Since the extra $Z_H$ boson can mix with the SM $Z$ boson or photon,
the EW precision observables may severely constrain our models.
For example, the mass mixing between the $Z$ and $Z_H$ bosons is strongly
restricted by the $\rho$ parameter. If the extra singlet field $\Phi$ 
does not exist, the mass of $Z_H$ must be of order of EW scale.
In this case, severe fine-tuning may be necessary in order that our models 
become consistent with the EW precision observables.

The basic ideas presented in this Letter could be readily applied 
to other cases, for example, to multi-Higgs doublet models 
in order to control the flavor problem by new gauge symmetries associated 
with Higgs fields.
In fact, the present authors recently constructed a new class of multi-Higgs 
doublet models which were designed to accommodate the top forward-backward
asymmetry at the Tevatron without conflict with the upper bounds on 
the same sign top pair production at the LHC \cite{Ko:2011vd,Ko:2011di,Ko:2012ud}. 
The model was based on the flavor dependent leptophobic $U(1)^{'}$ 
interactions, which can be also considered as Higgs flavor suggested 
in this Letter. 
By construction, those models had no serious flavor problems except for 
the large $u-t$ transitions. We can understand this 
surprising results observed in Refs.~\cite{Ko:2011vd,Ko:2011di,Ko:2012ud} 
using the new $U(1)_H$ symmetry  acting on (some of) the Higgs doublets 
as presented in this Letter.


{\it Note Added}

While we are revising this Letter, both the CMS and ATLAS Collaborations
have announced the discovery of a Higgs-like scalar boson 
with $\sim 125$ GeV mass~\cite{higgsdiscovery}.
The mass of this new Higgs-like scalar boson and its couplings 
to the SM particles will constrain our models as described in the text, 
when they are measured with reasonable accuracy in the future.

\section*{Acknowledgements}
\begin{acknowledgments}
We are grateful to P. Fayet, M. Drees, K.Y. Lee, Seung J. Lee, J.H. Park 
and J.L. Rosner for comments and suggestions.
Part of this work was done at the  Aspen Center for Physics in 2011, and 
partly supported by the National Science Foundation under Grant No. 1066293 
and the hospitality of the Aspen Center for Physics. 
This work is supported in part by National Research Foundation through 
Korea Neutrino Research Center at Seoul National University. 
This work is supported in part by Basic Science Research Program
through NRF 2011-0022996 (CY), by NRF Research Grant 
2012R1A2A1A01006053 (PK and CY), and by SRC program of NRF 
Grant No. 20120001176 through 
Korea Neutrino Research Center at Seoul National University (PK).
\end{acknowledgments}




\end{document}